\documentclass[prl,aps,twocolumn,superscriptaddress,floatfix,showpacs]{revtex4}
\usepackage{graphicx,rotating,subfigure,amsmath,amsfonts,amssymb,delarray}
\usepackage{dsfont}
\usepackage[T1]{fontenc}
\renewcommand{\vec}[1]{\boldsymbol #1}
\newcommand{\e}{\text{e}}

\def\l{\left}
\def\r{\right}
\def\12{\frac{1}{2}}

\newcommand{\be}{\begin{equation}}
\newcommand{\ee}{\end{equation}}
\newcommand{\bea}{\begin{eqnarray}}
\newcommand{\eea}{\end{eqnarray}}

\begin{document}
\bibliographystyle{apsrev}
\title{Spinons and helimagnons in the frustrated Heisenberg chain} 


\author{Jie Ren}
\affiliation{Department of Physics, Technical University Kaiserslautern,
D-67663 Kaiserslautern, Germany}

\author{Jesko Sirker} 
\affiliation{Department of Physics, Technical University Kaiserslautern,
D-67663 Kaiserslautern, Germany}
\affiliation{Research Center OPTIMAS,
  Technical University Kaiserslautern, D-67663 Kaiserslautern,
  Germany}
\date{\today}

\begin{abstract}
  We investigate the dynamical spin structure factor $S(q,\omega)$ for
  the Heisenberg chain with ferromagnetic nearest ($J_1<0$) and
  antiferromagnetic next-nearest ($J_2>0$) neighbor exchange using
  bosonization and a time-dependent density-matrix renormalization
  group algorithm. For $|J_1|\ll J_2$ and low energies we analytically
  find and numerically confirm two spinon branches with different
  velocities and different spectral weights. Following the evolution
  of $S(q,\omega)$ with decreasing $J_1/J_2$ we find that helimagnons
  develop at high energies just before entering the ferromagnetic
  phase.  Furthermore, we show that a recent interpretation of neutron
  scattering data for LiCuVO$_4$ in terms of two weakly coupled
  antiferromagnetic chains ($|J_1|\ll J_2$) is not viable. We
  demonstrate that the data are instead fully consistent with a
  dominant ferromagnetic coupling, $J_1/J_2\approx -2$.
  \end{abstract}


\maketitle
Frustrated spin models often show complicated phase diagrams.
Apart from phases with conventional (quasi) long-range magnetic order,
also valence bond solid, chirally ordered, multipolar as well as spin
liquid phases are possible. In addition to identifying and studying
the different phases, also the quantum phase transitions between them
have attracted considerable interest
\cite{SenthilViswanath,SirkerKrivnov}. Numerical studies are hindered
by the so-called sign problem making quantum Monte-Carlo simulations
impracticable. Our knowledge about frustrated systems in dimensions
$d>2$ is therefore still limited \cite{YanHuse,SirkerZheng,JiangYao}.
In one dimension, on the other hand, bosonization techniques can be
used to analytically study the weakly frustrated case
\cite{WhiteAffleck96,NersesyanGogolin}.  Furthermore, with the
density-matrix renormalization group (DMRG) \cite{WhiteDMRG} a
powerful numerical method is available to calculate static properties
at zero \cite{WhiteAffleck96} and finite temperatures \cite{SirkerMF}.
Lately, it has been shown that it is a useful strategy to study
frustration in two or three dimensions by a successive coupling of
one-dimensional chains \cite{KohnoStarykh,YanHuse} thus further
motivating investigations of the static and dynamic properties of
frustrated spin chains.

The standard spin model to study frustration in one dimension is the
$J_1-J_2$ Heisenberg model
\begin{equation}
\label{Ham}
H=\sum_j \l[J_1\vec{S}_j\vec{S}_{j+1} + J_2\vec{S}_j\vec{S}_{j+2}\r]
\end{equation}
with antiferromagnetic coupling $J_2>0$. Here $\vec{S}_j$ is a
spin-$1/2$ operator acting at site $j$. This model has recently been
studied intensely for ferromagnetic coupling $J_1<0$
\cite{BursillGehring,HikiharaKecke,SudanLuescher,VekuaHonecker,HeidrichMeisnerHonecker},
driven by the discovery of edge-sharing cuprate chain compounds for
which Eq.~(\ref{Ham}) seems to be the minimal model.  These cuprates
show fascinating properties including multiferroicity
\cite{MasudaZheludev,ParkChoi,SekiYamasaki,DrechslerVolkova,SchrettleKrohns},
i.e., an intricate interplay between incommensurate magnetic and
ferroelectric order \cite{KatsuraNagaosa}. It is, however, so far
unclear if all of their magnetic properties can be reasonably well
described within the simple minimal model (\ref{Ham}). In one of the
best studied edge-sharing chain cuprates LiCuVO$_4$, for example,
susceptibility data seem to point to a dominant ferromagnetic coupling
$\alpha\equiv J_1/J_2 \approx -2$ \cite{BuettgenKrugvonNidda,SirkerMF}
while neutron scattering data have been interpreted in terms of two
weakly coupled antiferromagnetic chains $\alpha\approx -0.7$
\cite{EnderleFak}, a conclusion which has later been challenged
\cite{EnderleFakcomment,NishimotoDrechsler,EnderleFakreply}.

Apart from being relevant for the edge-sharing cuprate chains, the
{\it dynamical} properties of the $J_1-J_2$ chain are also of
fundamental interest. For $\alpha =0$ the model consists of two
decoupled antiferromagnetic chains whose elementary gapless
excitations are spinons. Introducing a small coupling $J_1<0$ between
the chains leads classically to a spiral magnetic order while
bosonization predicts incommensurate spin correlations and an
exponentially small gap in the quantum $S=1/2$ case
\cite{NersesyanGogolin}. At $\alpha_c=-4$ there is an unusual quantum
critical point 
\cite{SirkerKrivnov} separating the incommensurate from a
ferromagnetic phase. Changing the frustration ratio $\alpha\leq 0$
thus turns antiferromagnetic spinons through an incommensurate phase
into ferromagnetic magnons.

In this letter we present a systematic study of the dynamical spin
structure factor 
\begin{equation}
\label{Sqw}
S(q,\omega)=\frac{1}{N}\sum_{j,j'} \e^{-iq(j-j')}\int dt \,\e^{i\omega t} \langle S^z_j(0)S^z_{j'}(t)\rangle
\end{equation}
of the $J_1$-$J_2$ Heisenberg model (\ref{Ham})
from the limit $\alpha=0$ of decoupled antiferromagnetic Heisenberg
chains, across the quantum critical point $\alpha_c$, into the
ferromagnetic phase, $\alpha <-4$. Note that due to $SU(2)$ symmetry
it is sufficient to consider the longitudinal correlation function in
(\ref{Sqw}). For $|\alpha|\ll 1$ we compare our data with results
obtained by bosonization while for $\alpha\sim -4$ we compare with
spin wave theory. 
Finally, we present a comparison of our results with recent neutron
scattering data for the multiferroic cuprate LiCuVO$_4$
\cite{EnderleFak}.

We start by considering the weak coupling limit $|\alpha|\ll 1$ by
bosonization. On each of the two antiferromagnetically coupled
sublattices we write
\begin{eqnarray}
\label{Bos1}
S^z_{2j(+1)}&=&\sqrt\frac{K_{1,2}}{\pi}\partial_x\phi_{1,2}+(-1)^j \mbox{const}\cos{\sqrt{4\pi K_{1,2}}\phi_{1,2}}\nonumber \\
S^+_{2j(+1)}&\propto & \e^{i\sqrt{\frac{\pi}{K}}\theta_{1,2}}\left[(-1)^j+\cos\sqrt{4\pi K_{1,2}}\phi_{1,2}\right] .
\end{eqnarray}
Here $\phi_{1(2)}$ are bosonic fields obeying the standard commutation
rules
$[\phi_\alpha(x),\partial_{x}\theta_{\alpha'}(x')]=i\delta_{\alpha,\alpha'}\delta(x-x')$
and $K_{1,2}$ are the Luttinger parameters. Let us first consider the
free fermion case with $J_2\vec{S}_i\vec{S}_j\to
\frac{J_2}{2}(S^+_iS^-_j +h.c.)$. Ignoring irrelevant terms,
bosonization leads to 
\begin{equation}
\label{Bos2}
H=\frac{1}{2}\sum_\alpha\int\, dx v_\alpha \left\{(\partial_x\phi_\alpha)^2+(\partial_x\theta_\alpha)^2\right\}
\end{equation}
for each of the chains $\alpha=1,2$. In this case $K_1=K_2=1$ and
$v_{1,2} =v_F$ with $v_F=2J_2$ in units of the lattice constant due to
a doubling of the unit cell. Apart from irrelevant terms, the
interchain coupling $J_1$ introduces a density-density type
interaction
\begin{equation}
\label{Bos3}
H_{d-d}=\frac{2J_1}{\pi}\int dx\,\partial_x\phi_1\partial_x\phi_2 \, .
\end{equation}
We can absorb this term into the Gaussian part (\ref{Bos2}) by
defining the new fields $\phi_{\pm}=(\phi_1\pm\phi_2)/\sqrt{2}$ and
$\theta_{\pm}=(\theta_1\pm\theta_2)/\sqrt{2}$. The Hamiltonian
(\ref{Bos2}) stays invariant under this transformation with
$\alpha=1,2$ being replaced by $\alpha=+,-$ but the velocities are
renormalized with $v_{1,2}\to v_\pm=v_F\pm J_1/\pi$.

In the isotropic Heisenberg case for $\alpha=0$ the low-energy
properties are still described by (\ref{Bos2}) but with
$v_{1,2}=J_2\pi$ and $K_{1,2}=1/2$ as known from Bethe ansatz. To keep
the $SU(2)$ symmetry intact, we now use non-Abelian instead of the
Abelian bosonization (\ref{Bos1}) for the interchain interaction
$J_1$. This leads to various marginal terms
\cite{NersesyanGogolin,AllenSenechal} and, in particular, a term which
can again be expressed as in (\ref{Bos3}) but with a different
prefactor $2J_1/\pi\to 3J_1/\pi$ leading to a velocity renormalization
\begin{equation}
\label{veloc}
v_{1,2}=\pi J_2 \to v_\pm =\pi J_2 (1\pm 3\alpha/2\pi^2) \, .
\end{equation}
The other marginal terms induced by the coupling $J_1$ have been
studied by renormalization group methods in \cite{NersesyanGogolin}
and seem to lead to an exponentially small gap. Numerically, however,
no gap in the parameter regime $-4<\alpha<0$ has been confirmed yet
and we will therefore neglect these 
terms for the moment.

At weak coupling $J_1$, the structure factor $S(q,\omega)$ will have
low-energy contributions at $q\sim 0,\pi/2,\pi$. We now calculate
these contributions in the isotropic case using the free boson
Hamiltonian (\ref{Bos2}) with $\alpha=+,-$ and velocities $v_\pm$
given by Eq.~(\ref{veloc}). Using the identity
(\ref{Bos1})---rewritten in terms of 
$\phi_\pm =(\phi_1\pm\phi_2)/\sqrt{2}$
---and the standard result for the boson propagator we obtain
\begin{eqnarray}
\label{Sq0w}
&& S(q\gtrsim 0[\pi-q\lesssim\pi],\omega)\\
&\approx& Kq (1\pm\cos q)\delta(\omega-v_{+}q) + Kq(1\mp\cos q)\delta(\omega-v_{-}q)\nonumber \\
&\approx & 2Kq \delta(\omega-v_{+[-]}q) +K\frac{q^3}{2}\delta(\omega-v_{-[+]}q) \,. \nonumber
\end{eqnarray}
The singularity at $\omega=v_+ q$ dominates $S(q\sim 0,\omega)$
while the singularity at $\omega=v_- q$ is dominant for $q\sim\pi$.
The other contribution is in both cases surpressed by a factor
$q^2/4$. Note that $v_+<v_-$ because $J_1<0$. Irrelevant band
curvature and interaction terms will lead to a finite linewidth
\cite{PereiraSirker,PereiraSirkerJSTAT}, an aspect which is beyond the
scope of the present study. For $|k|=|\pi/2-q|\ll 1$ and $v_+,v_-$
sufficiently different we find
\begin{eqnarray}
\label{Sqpi2w}
&&S(k,\omega)\propto \theta(\omega-v_+|k|) \\
&\times&\left\{(\omega^2-v_+^2k^2)^{K_- +\frac{K_+}{2}-1}+ |\omega^2-v_-^2k^2|^{K_+ +\frac{K_-}{2}-1}\right\}\, . \nonumber
\end{eqnarray}
Here $K_\pm\approx 1/2$ are the Luttinger parameters for the two modes
at weak coupling $J_1$. Note that the divergencies in (\ref{Sqpi2w})
are weaker than the square-root divergence for the pure isotropic
Heisenberg chain \cite{Schulz86}.  Taking the additional marginal
current-current interactions into account perturbatively, we find
furthermore an asymmetry $S(\pi/2-q,\omega)> S(\pi/2+q,\omega)$ with
$q>0$ small.

\begin{figure*}[!ht]
\begin{center}
\includegraphics*[angle=-90,width=0.98\textwidth]{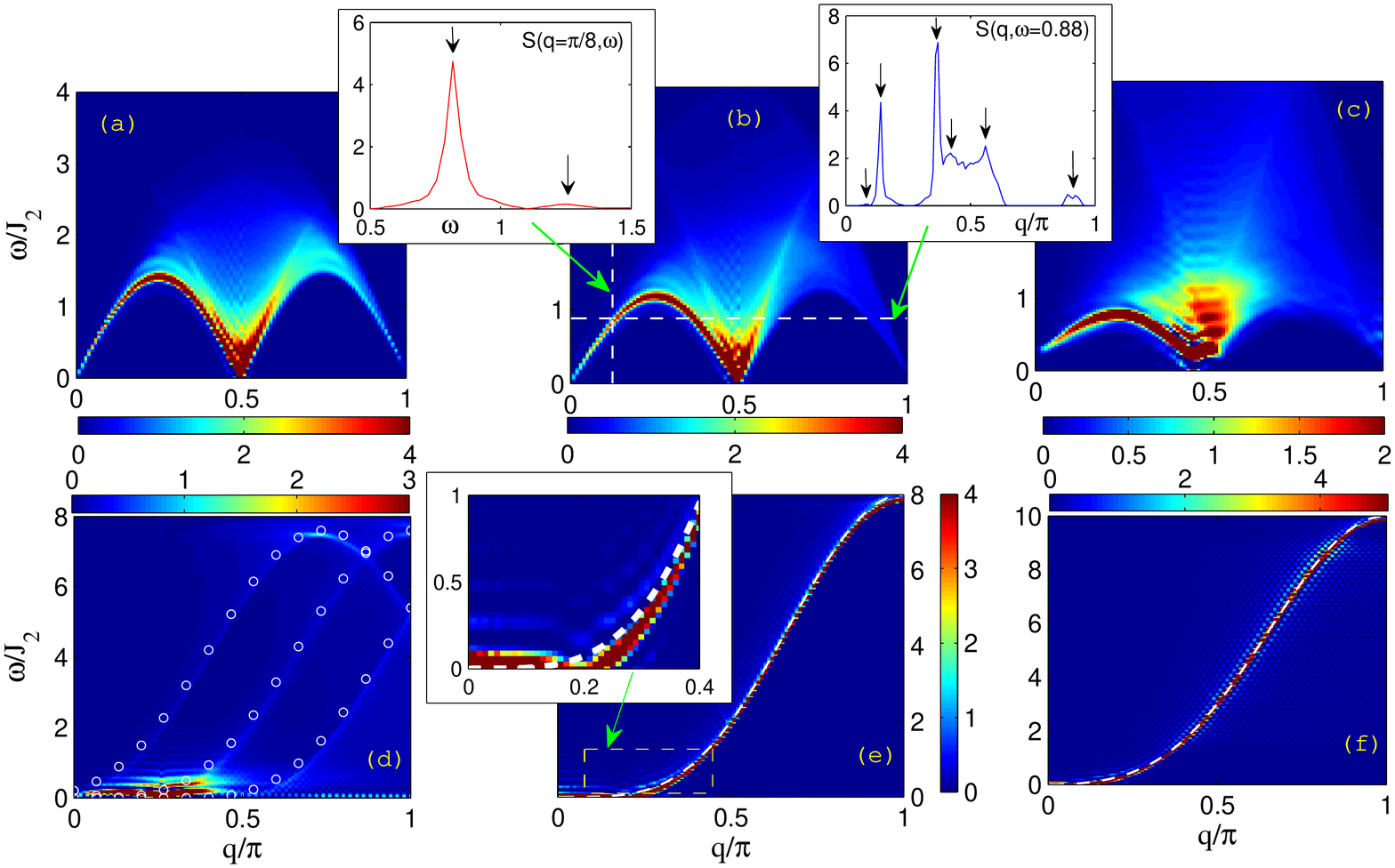}
\end{center}
\caption{(Color online) $S(q,\omega)$ for frustrations (a)
  $\alpha=-1/2$, (b) $\alpha=-1$, (c) $\alpha=-2$, (d) $\alpha=-3.5$,
  (e) at the quantum critical point $\alpha=-4$, and (f) in the
  ferromagnetic regime $\alpha=-5$. The insets to (b) show a constant
  $q$ and constant $\omega$ scan with arrows denoting spinon modes
  discussed in the text. The circles/dashed lines in (d-f) denote the
  magnon dispersions, Eq.~(\ref{helimagnon}).}
\label{Fig1}
\end{figure*}
To test the analytic predictions in the weak coupling limit and extend
our study to the experimentally relevant regime of strong interchain
coupling, we now turn to a numerical calculation of the dynamical spin
structure factor. We use an adaptive time-dependent density-matrix
renormalization group (DMRG) algorithm with a second order
Trotter-Suzuki decomposition of the time evolution operator
\cite{FeiguinWhite}. As time step we choose $J_2\delta t=0.1$.
We present results for an open chain with $N=400$ sites with $400$
states kept in the adaptive Hilbert space. In order to perform the
Fourier transform in time in Eq.~(\ref{Sqw}) one has to deal with the
problem that numerical data are only available for a finite time
interval $t\in [0,t_{\rm max}]$. The maximal simulation time, $t_{\rm
  max}\sim 40J_2$, up to which our numerical results are reliable has
been estimated by keeping track of the discarded weight and by
comparing with exact results for the $XX$ model and Bethe ansatz
results for the isotropic Heisenberg chain \cite{CauxMaillet2}. We
calculate the spin correlations $\langle S^z_{N/2}(0)S^z_{N/2\pm
  j}(t)\rangle$. Since two-point spin correlations are negligible for
distances much larger than $v_{\pm}t$, we obtain results almost
unaffected by the boundaries in the accessible time interval. These
data are then extended in time using linear prediction
\cite{Yule,BarthelSchollwockWhite} leading to a smooth exponentially
decaying extrapolation of the data.  We want to stress that linear
prediction does not allow to obtain reliable results for $t>t_{\rm
  max}$ but rather represents a smooth cutoff which does not affect
the data for $t<t_{\rm max}$.

\begin{figure}[h!]
\begin{center}
\includegraphics*[width=0.99\columnwidth]{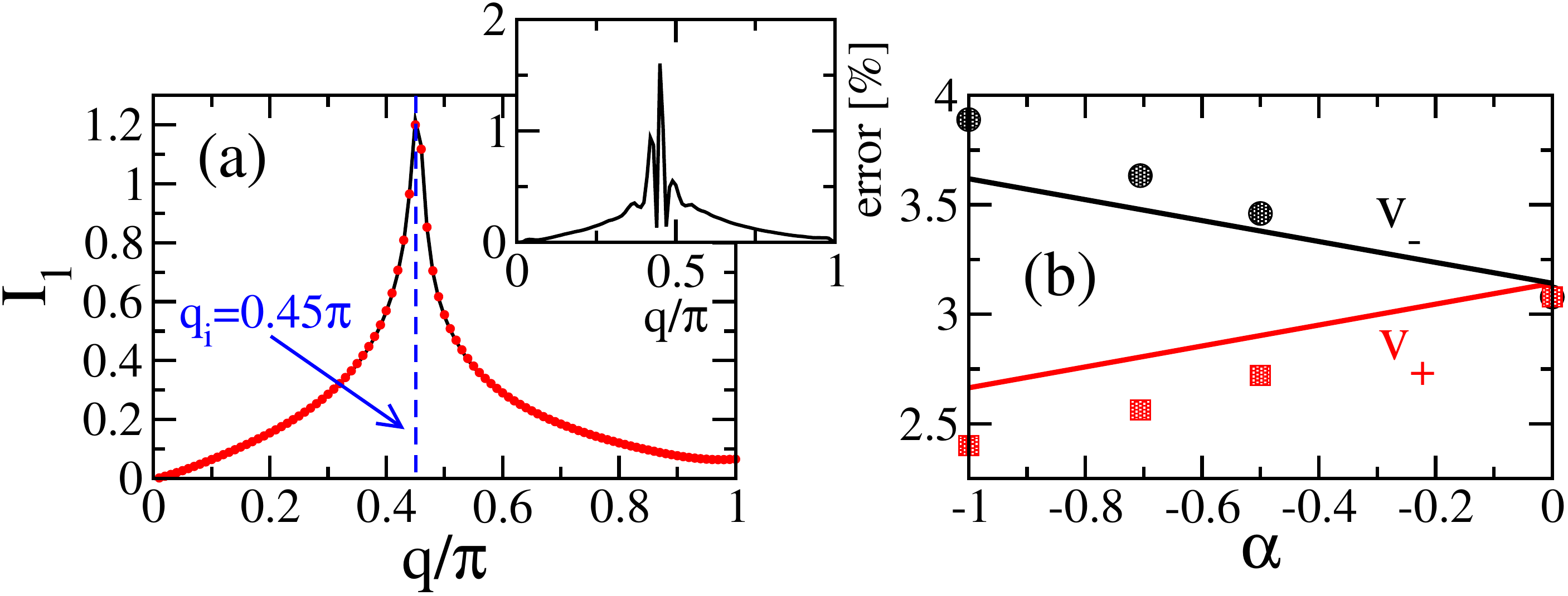}
\end{center}
\caption{(Color online) (a) Sum rule $I_1$ for $\alpha=-2$ from
  frequency-integrating the numerical data (symbols) compared to a
  static DMRG calculation (line). The inset shows the absolute error
  of the frequency-integrated data. (b) Velocities of the elementary
  excitations $v_\pm$.  The symbols denote the values extracted from
  the numerical data for $S(q,\omega)$ at $q\sim 0,\pi$ the lines the
  prediction from bosonization, Eq.~(\ref{veloc}).}
\label{Fig2}
\end{figure}

In Fig.~\ref{Fig1} we show our results for $S(q,\omega)$ for various
frustrations.
A further check of the quality of the numerical data is obtained by
considering the sum rules $I_1=\int \frac{d\omega}{2\pi}
S(q,\omega)=\langle S^z_q S^z_{-q}\rangle$ where $S^z_q=N^{-1/2}\sum_j
\e^{-iqj}S^z_j$ and $I_2=\sum_q I_1 =1/4$. For all frustrations shown
we find that the sum rules are fulfilled with an absolute error of at
most $2\%$. As an example, we show results for $I_1$ at $\alpha=-2$ in
Fig.~\ref{Fig2}(a).

For weak frustrations $\alpha=-1/2,-1$ (Fig.~\ref{Fig1}(a,b)) two
excitations with different velocities $v_\pm$ are clearly visible near
$q\sim \pi/2$. As expected from bosonization, the smaller velocity
$v_+$ agrees with that of the dominant excitation at $q\sim 0$ while
the one at $q\sim\pi$ has velocity $v_-$. The velocities extracted
from the numerical data also agree fairly well with the prediction
from bosonization, see Fig.~\ref{Fig2}(b).
A closer inspection of the data for $\alpha=-1$ also reveals the
second mode with velocity $v_-$ at $q\sim 0$ with a weight suppressed
by approximately $q^2/4$ relative to the dominant mode, see inset of
Fig.~\ref{Fig1}(b).

At larger frustrations the weak coupling picture from bosonization
clearly breaks down. For $\alpha=-2,-3.5$ shown in Fig.~\ref{Fig1}(c)
and (d) the low energy spectral weight is concentrated at an
incommensurate wave vector $q_i$. Classically, frustration leads to
the formation of a helical state with a pitch vector
$Q=\arccos(|\alpha|/4)$. For the quantum model, incommensurate spin
correlations have been shown to occur with wave vectors $q_i$ which
approach $\pi/2$ with increasing $\alpha$ much faster than in the
classical case \cite{BursillGehring,SirkerMF} in full agreement with
our dynamical data. For $\alpha=-3.5$ we observe---in addition to the
low-energy spectral weight near $q_i\approx 0.26\pi$---the development
of three magnon-like dispersions at higher energies. For the
classically expected state with long-range spiral order at wave vector
$Q$, spin-waves have the dispersion
\begin{equation}
\label{helimagnon}
\epsilon_q=\sqrt{A_q^2-B_q^2}
\end{equation}
with $A_q=-J_Q+J_q/2+(J_{q+Q}+J_{q-Q})/4$, $B_q=-J_q/2 +
(J_{q+Q}+J_{q-Q})/4$ where $J_q=J_1\cos q +J_2\cos 2q$
\cite{Nagamiya,ZhitomirskyZaliznyak}. 
In the quantum model, long-range order is destroyed. However, close to
the quantum critical point $\alpha_c=-4$ where the model starts to
develop a long-ranged ordered ferromagnetic state, the correlation
length will become large so that the excitations at high energy remain
helimagnon-like. As shown in Fig.~\ref{Fig1}(d) the helimagnon
dispersion (\ref{helimagnon}) does indeed describe the high-energy
modes very well with a renormalized effective $\alpha=-3.85$
corresponding to a pitch vector $Q=\arccos(|\alpha|/4)\approx 0.27$.
The three modes \cite{Nagamiya} are then given by $\epsilon_q$ and
$\epsilon_{q\pm q_i}$ where $q_i$ is the incommensurate wave vector of
the {\it quantum} model.

For $\alpha<-4$ the ground state is a simple ferromagnet, $Q=0$, and
(\ref{helimagnon}) reduces to the magnon dispersion
$\epsilon_q=-J_2\left[(1-\cos 2q)+\alpha(1-\cos q)\right]$ which is in
excellent agreement with the numerical data as shown in
Fig.~\ref{Fig1}(f). At the quantum critical point, $\alpha_c=-4$, the
magnon dispersion becomes quartic, $\epsilon_q=J_2q^4/2$ at small $q$.
As already noticed in \cite{SirkerKrivnov} spin-wave theory does not
describe the quantum critical point correctly at low energies due to
the degeneracy of the ferromagnetic with valence bond solid states
\cite{HamadaKane}.  This is confirmed by our data (see inset of
Fig.~\ref{Fig1}(e)) showing that the low-energy spectral weight does
not follow the magnon dispersion while the agreement is good at higher
energies.

Let us finally discuss our results in the light of recent neutron
scattering experiments on LiCuVO$_4$ \cite{EnderleFak} and LiCuSbO$_4$
\cite{DuttonKumar}.
\begin{figure}[h!]
\begin{center}
\includegraphics*[width=0.99\columnwidth]{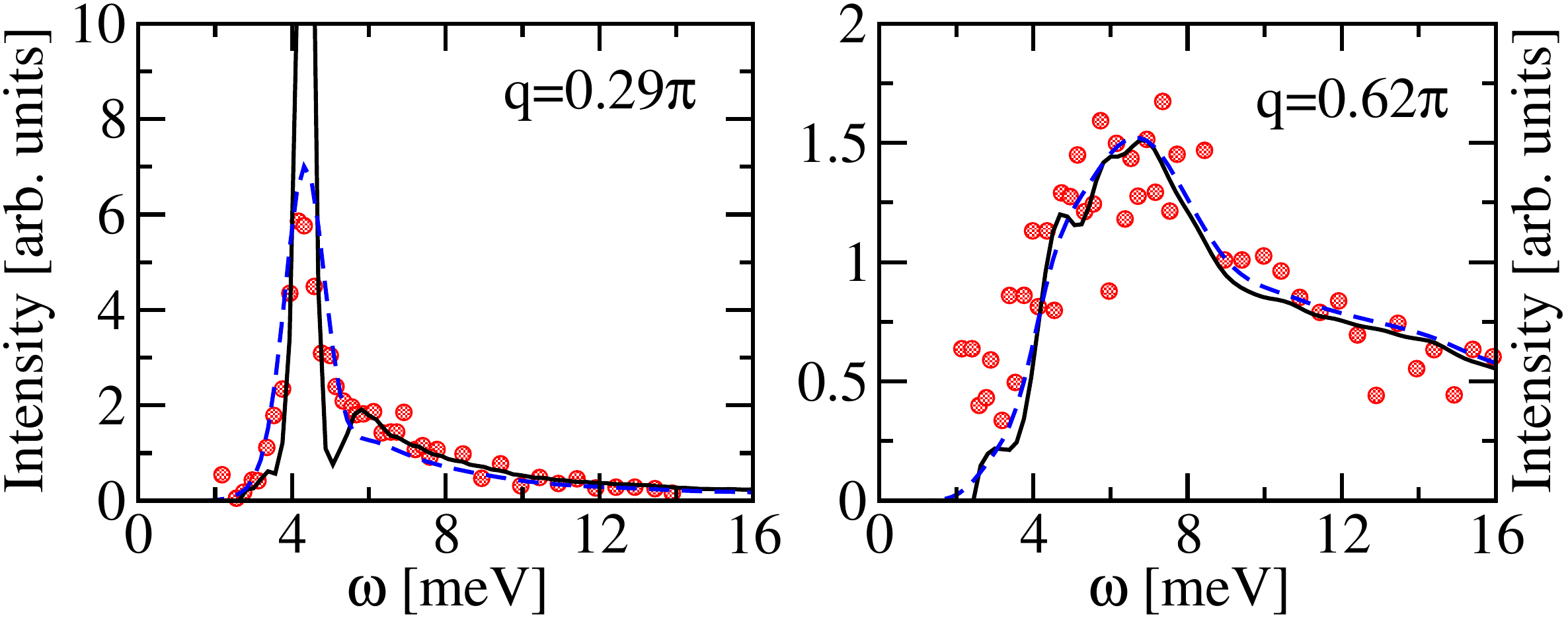}
\end{center}
\caption{(Color online) Neutron scattering data taken from Fig.~3 of
  Ref.~\cite{EnderleFak} (circles) compared to the DMRG data with
  $J_2=6$ meV and $\alpha=-2$ (solid lines). The dashed lines are
  the same DMRG spectra broadened by convoluting with a Gaussian with
  a FWHM of $1.1$ meV. The height of the theoretical spectra has been
  scaled to fit the data best.}
\label{Fig3}
\end{figure}
The experimental data for LiCuVO$_4$ have been interpreted in terms of
the $J_1$-$J_2$ Heisenberg model with $|J_1|<J_2$. However, none of
the features predicted by bosonization in this limit and visible in
Fig.~\ref{Fig1}(a,b) have been observed. The neutron data in Fig.~2 of
\cite{EnderleFak} look instead remarkably similar to our numerical
results for $\alpha=-2$. Comparing to the numerically calculated
structure factor one has to keep in mind that the neutron scattering
data are slightly supressed at higher momenta due to the
$q$-dependence of the atomic form factor for Cu$^{2+}$ ions and the
whole spectrum is broadened due to a finite $q$, $\omega$ resolution.
However, this does not affect the qualitative features of the
spectrum. In fact, we can even obtain a quantitatively satisfying
description of the data, see Fig.~\ref{Fig3}. For small
momenta
the agreement becomes excellent when convoluting the numerical data
with a Gaussian to take the finite instrumental resolution into
account. A frustration $\alpha\sim -2$ implies an incommensurate wave
vector $q_i\approx 0.45\pi$, see Fig.~\ref{Fig2}(a), consistent with
the neutron data and susceptibility measurements
\cite{BuettgenKrugvonNidda,SirkerMF}. For such a strong frustration
the explanation of the spectral weight 
at high energies in terms of multispinon excitations of the
antiferromagnetic chains offered in Ref.~\cite{EnderleFak} is not
viable. Instead, this weight is related to the incommensurate spin
correlations in this material caused by the dominant ferromagnetic
coupling between the chains. Let us also briefly comment on very
recent neutron scattering results on powder samples of LiCuSbO$_4$
\cite{DuttonKumar}. The powder averaging prevents a detailed analysis,
however, the concentration of spectral weight at the incommensurate
wave vector $q_i\approx 0.47\pi$ points again to a strong frustration
$\alpha\approx -2$ which is fully consistent with an analysis of
susceptibility and specific heat data \cite{DuttonKumar}.

\acknowledgments The authors thank J.-S. Caux for sending us his data
for the Heisenberg chain. J.S. acknowledges support by the DFG via the
SFB/TR 49 and by the graduate school of excellence MAINZ and J. R. by
the National Natural Science Foundation of China (NO.11104021).


\end{document}